\documentclass[aps,floatfix,twocolumn,showpacs,pra]{revtex4}
\usepackage{graphicx,bm}
\usepackage{amsmath}
\usepackage{amssymb}
\newcommand{\be}{\begin{equation}}
\newcommand{\ee}{\end{equation}}
\newcommand{\bea}{\begin{eqnarray}}
\newcommand{\eea}{\end{eqnarray}}
\newcommand{\rr}{\mathbf{r}}
\begin{document}
\title{Particle number fluctuations in a cloven trapped Bose gas at finite temperature}
\author{A. Sinatra}
\affiliation{Laboratoire Kastler Brossel,
Ecole Normale Sup\'erieure, UPMC and CNRS,
24 rue Lhomond, 75231 Paris Cedex 05, France}
\author{Y. Castin}
\affiliation{Laboratoire Kastler Brossel,
Ecole Normale Sup\'erieure, UPMC and CNRS,
24 rue Lhomond, 75231 Paris Cedex 05, France}
\author{Yun Li}
\affiliation{State Key Laboratory of Precision Spectroscopy, Department of Physics,
East China Normal University, Shanghai 200062, China}

\begin{abstract}
We study fluctuations in the atom number difference between two halves of a harmonically trapped Bose gas in three dimensions.
We solve the problem analytically for non interacting atoms. In the interacting case we find an analytical 
solution in the Thomas-Fermi and high temperature limit in good agreement with classical field simulations. In the large system size limit,
fluctuations in the number difference are maximal for a temperature $T\simeq 0.7 \,T_c$ where $T_c$ is the critical temperature,
independently of the trap anisotropy. The occurrence of this maximum is due to an interference effect between the condensate and the non-condensed field.
\end{abstract}


\maketitle

\section{Introduction}

Fluctuations appearing when counting the atoms in a given sub-volume of a quantum system, are a fundamental 
feature determined by the interplay between the atomic interactions and quantum statistics.  
They can be used to investigate many-body properties of the
system and in particular non-local properties of the $g^{(2)}$ pair-correlation function. 
These fluctuations were studied at zero temperature for quantum gases in different regimes and spatial dimensions in \cite{Molmer,Houches2003,VarennaYvan,Combescot,Brustein}. 
Sub-poissonian fluctuations appear for non-interacting fermions and for interacting bosons. A related issue in condensed matter physics
is that of partition noise in electron systems \cite{Levitov}.

In cold atoms experiments it is now possible to directly measure the fluctuations in atom number within a given region, as done for example in  \cite{Bouchoule} for a
quasi one-dimensional system. However finite temperature plays a major role in experiments.
Very recently, experiments were done on an atom chip where a cold gas of Rb atoms, initially trapped in an single harmonic potential well, is split
into two parts by raising a potential barrier. Accurate statistics of the particle number difference between the left and right wells $N_L-N_R$ is then performed
in the modified potential. 
By varying the initial temperature of the sample across the transition for Bose-Einstein condensation, they observe a marked peak in the fluctuations 
of the particle number difference below the transition temperature $T_c$ while
shot noise fluctuations are recovered for $T>T_c$. For $T \ll T_c$ they finally get sub-shot noise fluctuations due to the repulsive interactions between
atoms \cite{Kenneth}. Here we show that the peak of fluctuations for $T<T_c$ is in fact a general feature already appearing in a single harmonic
well if we look at the fluctuations of the atom number difference $N_L-N_R$ between the left half and the right half of the trap along one axis. We show
that, contrarily to what happens to the total number fluctuations, fluctuations in the particle number difference $N_L-N_R$ 
can be computed within the grand canonical ensemble without any pathology. 
In the first part of the paper we address the ideal gas case for which we find the complete analytical solution in the grand canonical ensemble.
We derive the asymptotic behaviors for $T \ll T_c$ and $T \gg T_c$ and
we explain the physical origin of the ``bump" in fluctuations of the particle 
number difference for $T<T_c$. In the second part of our paper we then address the interacting case. 

\section{Ideal gas: exact solution}
\begin{figure}[hob]
\centerline{\includegraphics[width=7cm,clip=]{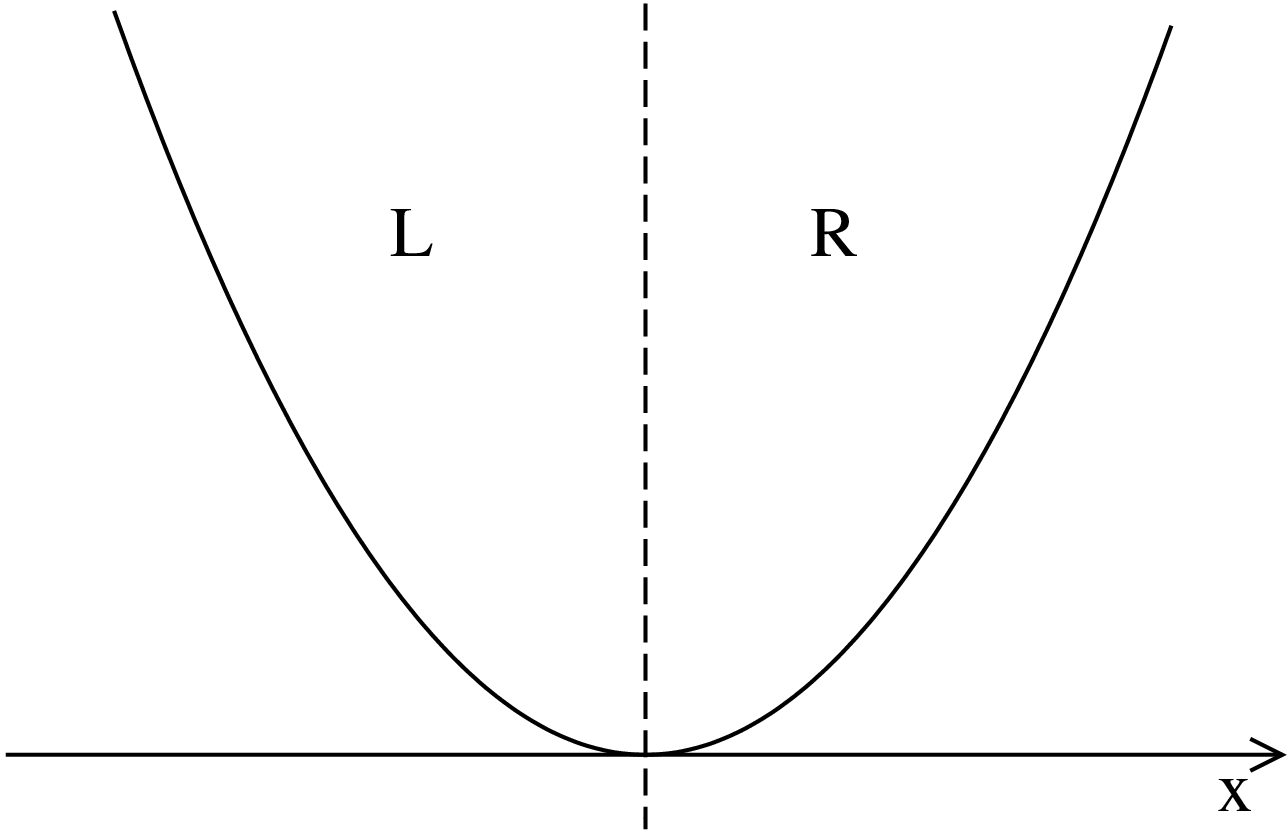}}
\caption{We consider fluctuations of the particle number 
difference $N_L-N_R$ between the left and the right halves 
of a three-dimensional harmonic potential. \label{fig:parab}}
\end{figure}

We consider an ideal gas of bosons in a three-dimensional harmonic potential. 
The signal we are interested in is the particle number difference $N_L-N_R$ 
between the left and right halves of the harmonic potential along one direction, 
as shown in Fig.\ref{fig:parab}. 
In terms of the atomic field operators:
\begin{equation}
N_L-N_R=\int_{\mathbf{r}\in L} {\psi}^\dagger \psi - \int_{\mathbf{r}\in R} {\psi}^\dagger \psi \,.
\end{equation}
Due to the symmetry of the problem, $N_L-N_R$ has a zero mean value.
It is convenient to express its variance in terms of the
unnormalized pair correlation function
\begin{equation}
g^{(2)}(\mathbf{r},\mathbf{r'})=\langle \psi^\dagger(\mathbf{r}) \psi^\dagger(\mathbf{r'}) \psi(\mathbf{r'}) \psi(\mathbf{r}) \rangle \,.
\end{equation}
Normally ordering the field operators with the help of the bosonic commutation
relation pulls out a term equal to the mean total number of particles:
\begin{eqnarray}
\label{eq:one}
\mbox{Var}(N_L-N_R) &=& \langle N \rangle +
2 \left[ \int_{\mathbf{r}\in L} \int_{\mathbf{r'}\in L} g^{(2)}(\mathbf{r},\mathbf{r'}) \right.
\nonumber \\
&-&  \left. \int_{\mathbf{r}\in L} \int_{\mathbf{r'}\in R}  g^{(2)}(\mathbf{r},\mathbf{r'}) \right] \,. \label{eq:prima}
\end{eqnarray}
We assume that the system is in thermal equilibrium in the grand canonical ensemble with $\beta=1/k_BT$ the inverse temperature and 
$\mu$ the chemical potential. Since the density operator is Gaussian, we can use 
Wick's theorem and express $g^{(2)}(\mathbf{r},\mathbf{r'})$ in terms
of the first-order coherence function  
$g^{(1)}(\mathbf{r},\mathbf{r'})
=\langle \psi^\dagger(\mathbf{r}) \psi(\mathbf{r'}) \rangle$:
\begin{equation}
g^{(2)}(\mathbf{r},\mathbf{r'})=g^{(1)}(\mathbf{r},\mathbf{r'})g^{(1)}(\mathbf{r'},\mathbf{r})+g^{(1)}(\mathbf{r},\mathbf{r})g^{(1)}(\mathbf{r'},\mathbf{r'}) \,.
\end{equation}
The $g^{(1)}(\mathbf{r},\mathbf{r'})$ is a matrix element of the one-body density operator 
\begin{equation}
g^{(1)}(\mathbf{r},\mathbf{r'})=\langle \mathbf{r'}| \frac{1}{z^{-1} e^{\beta h_1}-1}
|\mathbf{r} \rangle 
\end{equation}
where $h_1$ is the single particle Hamiltonian
\begin{equation}
h_1=\frac{\mathbf{p}^2}{2m}+ \sum_{\alpha=x,y,z} \frac{1}{2} m \omega_\alpha^2 r_\alpha^2 \,.
\end{equation}
To compute $g_1$, it is convenient to expand the one-body density operator
in powers of the fugacity $z=e^{\beta \mu}$ \cite{LesHouches}:
\begin{equation}
g^{(1)}(\mathbf{r},\mathbf{r'})=\langle \mathbf{r'}| \sum_{l=1}^{\infty} z^l e^{-l\beta h_1} |\mathbf{r} \rangle \,.
\end{equation}
On the other hand for a harmonic potential the matrix elements of $e^{-\beta h_1}$ are known \cite{Landau}. We then have:
\begin{eqnarray}
\label{eq:g1}
g^{(1)}(\mathbf{r},\mathbf{r'}) = \sum_{l=1}^{\infty} z^l  \left( \frac{m \bar{\omega}}{2\pi \hbar} \right)^{3/2}  
\prod_{\alpha=x,y,z} \left[ \sinh (l \eta_\alpha) \right]^{-1/2} &\times&  \nonumber  \\
 \exp \left\{ - \frac{m \omega_\alpha}{4\hbar} \left[ (r_\alpha+r'_\alpha)^2 \tanh \left(\frac{l\eta_\alpha}{2}\right) \right. \right. &+&  \nonumber \\
 \left. \left. (r_\alpha-r'_\alpha)^2 \coth \left(\frac{l\eta_\alpha}{2}\right) \right] \right\}  
\end{eqnarray}
where we introduced the geometric mean of the oscillation frequencies
$\bar{\omega}=(\omega_x \omega_y \omega_z)^{1/3}$ and $\eta_\alpha=\beta \hbar \omega_\alpha$.
It is convenient to renormalize the fugacity introducing $\tilde{z}=z \exp(-\sum_{\alpha} \eta_\alpha/2)$ that spans the interval $(0,1)$.
After some algebra \cite{Integrales}, 
the variance of $N_L-N_R$ is expressed as a double sum that we reorder as
\begin{equation}
\mbox{Var}(N_L-N_R) = \langle N \rangle  + \sum_{s=1}^{\infty}  c_s \tilde{z}^s \label{eq:result}
\end{equation}
with
\begin{multline}
c_s =   \displaystyle{\sum_{l=1}^{s-1}} \frac{
 1-\frac{4}{\pi} \arctan  \sqrt{\tanh (\frac{1}{2}l\eta_x) \tanh \left[ \frac{1}{2}(s-l)\eta_x \right] }}{\displaystyle{\prod_{\alpha=x,y,z}}
\left[ 1 - e^{-\eta_\alpha s} \right] },
\end{multline}
with $c_1=0$. Correspondingly, the mean atom number is expressed as
\begin{equation}
\langle N \rangle = \sum_{l=1}^{\infty} \tilde{z}^l 
\prod_\alpha \frac{1+\coth(l\eta_\alpha/2)}{2}.
\label{eq:nat}
\end{equation}
This constitutes our analytical solution 
of the problem in the grand canonical ensemble. 

\begin{figure}[hob]
\centerline{\includegraphics[width=7cm,clip=]{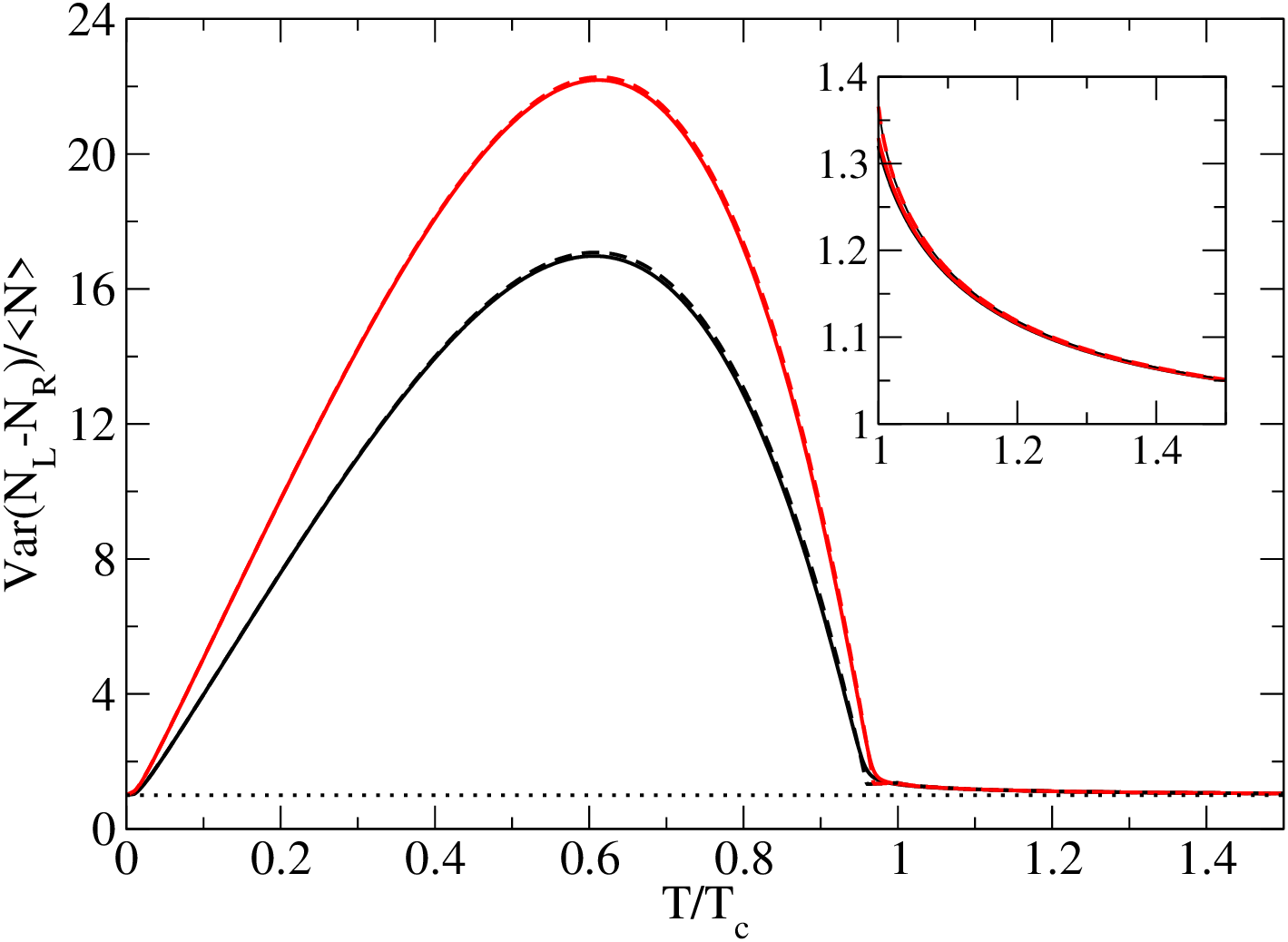}}
\caption{(Color online) Normalized variance of the particle number difference
$N_L-N_R$ as a function of temperature in a cigar-shaped trap
with $\omega_y=\omega_z=2 \omega_x$. The number of particles 
is $\langle N\rangle=6000$ (black lines) and $\langle N\rangle=13000$ (red lines).
The inset is a magnification of the $T>T_c$ region.
Solid lines: exact result (\ref{eq:result}) together with (\ref{eq:nat}). Dashed line for $T>T_c$: approximate result (\ref{eq:ht}) together
with (\ref{eq:nat_ht}).  Dashed line for $T<T_c$: approximate result 
resulting from the improved estimates
(\ref{eq:euler}) and (\ref{eq:ltb_nat}).  
The temperature $T$ is expressed in units of the critical temperature $T_c$ defined in Eq. (\ref{eq:Tc}). \label{fig:varn}}
\end{figure}

In practice, the forms (\ref{eq:result}) and (\ref{eq:nat}) are difficult to
handle in the degenerate regime, since the series converge very slowly
when $\tilde{z}\to 1$. A useful exact rewriting is obtained by pulling out 
the asymptotic behaviors of the summands.
For the signal we obtain the operational form
\begin{equation}
\mbox{Var}(N_L-N_R) = \langle N \rangle  + c_\infty \langle N_0 \rangle + \sum_{s=1}^{\infty} ( c_s -c_\infty)  \tilde{z}^s 
\label{eq:split}
\end{equation}
where
\begin{equation}
\label{eq:clim}
c_\infty = \lim_{s\to \infty} c_s = 2 \sum_{l=1}^\infty \left( 1 - \frac{4}{\pi} \arctan \sqrt{\tanh \frac{l \eta_x}{2} }\, \right) \,,
\end{equation}
and $\langle N_0 \rangle=\tilde{z}/(1-\tilde{z})$ is the mean number of condensate particles. The mean atom number is rewritten as
\begin{equation}
\label{eq:split_nat}
\langle N\rangle = \langle N_0\rangle + \sum_{l=1}^{\infty}
\tilde{z}^l \left[-1 + 
\prod_\alpha \frac{1+\coth(l\eta_\alpha/2)}{2}\right].
\end{equation}
In Fig. \ref{fig:varn} we show an example of fluctuations of the particle number
difference for realistic parameters of an atom-chip experiment.
It is apparent that fluctuations are weakly super-poissonian above $T_c$ and a marked peak of fluctuations occurs for $T<T_c$. 
In the following sections we perform some approximations or transformations
in order to obtain
explicit formulas and get some physical insight.

\section{Approximate formulas for $k_BT \gg \hbar \omega$}

In this section, we consider the limit of a large atom number 
and a high temperature $k_B T \gg \hbar \omega_\alpha$ for all $\alpha$.

{\it Non-condensed regime:}
Taking the limit $\eta_\alpha \ll 1$ in Eq. (\ref{eq:nat}) we get
\begin{equation}
\langle N \rangle \simeq \left( \frac{k_BT}{\hbar \bar{\omega}}\right)^3 g_3(\tilde{z})\, \label{eq:nat_ht} 
\end{equation}
where $g_\alpha(z)=\sum_{l=1}^\infty z^l/l^\alpha$ is the Bose function. 
From this equation we recover the usual definition of the
critical temperature $T_c$:
\begin{equation}
\label{eq:Tc}
k_B T_c = [N/\zeta(3)]^{1/3} \hbar \bar{\omega} \,,
\end{equation}
where $\zeta(3)=g_3(1)$  with $\zeta$ the Riemann function.
Taking the same limit in (\ref{eq:result}) gives
\begin{equation}
\mbox{Var}(N_L-N_R) \simeq \langle N \rangle \left[
1+ \frac{g_2(\tilde{z})-g_3(\tilde{z})}{\zeta(3)} \frac{T^3}{T_c^3}
\right]. \label{eq:ht}
\end{equation}
At $T=T_c$ this leads to 
\begin{equation}
\mbox{Var}(N_L-N_R)(T_c) \simeq N\frac{\zeta(2)}{\zeta(3)}\simeq 1.37 N\,,
\end{equation}
showing that the non-condensed gas is weakly super-poissonian,
as already observed in Fig.\ref{fig:varn}.

Alternatively, one may directly take the limit $\eta_\alpha\to 0$ 
in Eq.(\ref{eq:g1}), yielding
\begin{multline}
g^{(1)}(\mathbf{r},\mathbf{r'}) \simeq \sum_{l=1}^{\infty} \frac{\tilde{z}^l}{l^{3/2} \lambda_{dB}^3}    
\prod_{\alpha=x,y,z}  \\
\exp \left\{ - \frac{l m \omega_\alpha^2}{2 k_BT} \left( \frac{r_\alpha+r'_\alpha}{2} \right)^2 
  + \frac{\pi}{l \lambda_{dB}^2} (r_\alpha-r'_\alpha)^2  \right\} . \label{eq:g1lda}
\end{multline}
This semiclassical approximation coincides with the widely used local density
approximation, and allows to recover (\ref{eq:nat_ht}) and (\ref{eq:ht}).

{\it Bose-condensed regime:} In this regime $\tilde{z}\to 1$ 
so we use the splittings (\ref{eq:split}) and (\ref{eq:split_nat}). 
Setting $\tilde{z}=1$ and taking the limit 
$\eta_\alpha\to 0$ in each term of the sum over
$l$ in (\ref{eq:split_nat}), we obtain the usual condensate fraction
\begin{equation}
\frac{\langle N_0\rangle}{\langle N\rangle} \simeq  1 - \frac{T^3}{T_c^3}.
\label{eq:lt_nat}
\end{equation}
The same procedure may be applied to the sum over $s$ in (\ref{eq:split}).
The calculation of the small-$\eta$ limit of $c_\infty$ requires a different
technique: Contrarily to the previous cases, the sum in (\ref{eq:clim})
is not dominated by values of the summation index $\ll 1/\eta_\alpha$
and explores high values of $l\sim 1/\eta_x$. 
As a remarkable consequence, the local-density approximation fails in this case
\cite{lda}.
We find that one rather has to replace the sum over $l$
by an integral in (\ref{eq:clim}):
\begin{equation}
\label{eq:climint}
c_\infty \simeq \int_0^{+\infty} dl\, f(l) = \frac{2\ln 2}{\eta_x} 
\end{equation}
with $f(x)=2-(8/\pi)\arctan\sqrt{\tanh(x\eta_x/2)}$.
This leads to the simple formula for $T<T_c$:
\begin{multline}
\label{eq:lt}
\mbox{Var}(N_L-N_R) \simeq \langle N \rangle\left[1 + 
\frac{\zeta(2)-\zeta(3)}{\zeta(3)} \frac{T^3}{T_c^3} \right.\\
\left.+2\ln 2 \left( \frac{k_BT_c}{\hbar \omega_x} \right)
\left(1 - \frac{T^3}{T_c^3}\right)\, \frac{T}{T_c}
\right].
\end{multline}
The second line of Eq.(\ref{eq:lt}) is a new contribution involving 
the macroscopic value of $\langle N_0\rangle$ below the critical temperature.
Since $k_B T_c \gg \hbar \omega_x$ here, it is the dominant contribution
to the fluctuations of the particle number difference. 
It clearly leads to the occurrence
of a maximum of these fluctuations, at a temperature which remarkably
is independent of the trap anisotropy:
\begin{equation}
\left( \frac{T}{T_c} \right)_{\rm max} \simeq 2^{-2/3} \simeq 0.63 \,.
\label{eq:pos_max}
\end{equation}
The corresponding variance is strongly super-poissonian in the large atom-number
limit: \begin{equation} \left[ \mbox{Var}(N_L-N_R)\right]_{\rm max} \simeq \langle N \rangle \left[ 1+
\frac{3}{4} \ln 2 \left( \frac{2 \langle N \rangle}{\zeta(3)} \right)^{1/3}
\frac{\bar{\omega}}{\omega_x} \right] \label{eq:max}.
\end{equation}

For the parameters of the upper curve in
Fig.~\ref{fig:varn}, these approximate
formulas lead to a maximal variance over $\langle N\rangle$ 
equal to $\simeq 27.5$, whereas the exact result is $\simeq 22.2$,
located at $T/T_c\simeq 0.61$.
We thus see that finite size corrections remain important 
even for the large atom number $\langle N\rangle=13000$. 
Fortunately, it is straightforward to calculate
the next order correction. For the condensate atom number we simply
expand the summand in (\ref{eq:split_nat}) 
up to order $\eta_\alpha^2$, and we recover the known result \cite{Qui_citer}:
\begin{equation}
\frac{\langle N_0\rangle}{\langle N\rangle} \simeq  1 - \frac{T^3}{T_c^3}
-\frac{T^2}{T_c^2} \frac{3\zeta(2)}{2\zeta(3)} \frac{\hbar\omega_m}{k_B T_c}
\label{eq:ltb_nat}
\end{equation}
with the arithmetic mean 
$\omega_m=\sum_\alpha \omega_\alpha/3$.
For $c_\infty$ we use the Euler-Mac Laurin summation formula,
applied to the previously defined function $f(x)$ over the interval
$(1,+\infty)$, and we obtain
\begin{equation}
\label{eq:euler}
c_\infty = \frac{2\ln 2}{\eta_x} -1 + A \eta_x^{1/2} + O(\eta_x^{3/2})
\end{equation}
with $A = - 2^{5/2} \zeta(-1/2)/\pi \simeq 0.374$ \cite{details}.
Using these more accurate formulas for $\langle N_0\rangle$ and $c_\infty$
in the second term of the right-hand side of (\ref{eq:split}) leads
to an excellent agreement with the exact result, see the dashed
lines in Fig.~\ref{fig:varn} practically indistinguishable from the solid lines.
Note that the effect of the $-1$ correction in (\ref{eq:euler}) is to change 
the shot noise term $1$ in the square brackets (\ref{eq:lt}) and (\ref{eq:max}) into
$1-\langle N_0\rangle/\langle N \rangle$.

\section{A physical analysis singling out the condensate mode}
To investigate the contribution of different physical effects on our 
observable, it is convenient to go back to the expression (\ref{eq:one}) 
and split the field operator into the condensate and
the non-condensed part:
\begin{equation}
\psi(\mathbf{r})=\phi(\mathbf{r})a_0 + \delta \psi(\mathbf{r}) \,, \label{eq:sepa}
\end{equation}
where $\phi(\mathbf{r})$ is the ground mode wavefunction of the harmonic potential.
The pair correlation function $g^{(2)}$ is then expressed as the sum of three
contributions, $g^{(2)}(\mathbf{r},\mathbf{r'})=
g_I^{(2)}+g_{II}^{(2)}+g_{III}^{(2)}$,
sorted by increasing powers of $\delta\psi$:
\begin{eqnarray}
g_I^{(2)}&=& 
\phi^2(\mathbf{r})\phi^2(\mathbf{r'})
\langle a_0^\dagger a_0^\dagger a_0 a_0 \rangle 
\\
g_{II}^{(2)} &=& \left[ \phi(\mathbf{r})\phi(\mathbf{r'})
\langle a_0^\dagger a_0 \, \delta \psi^\dagger(\mathbf{r'})\delta \psi(\mathbf{r})     \rangle +
\mathbf{r} \leftrightarrow \mathbf{r'} \right]
\nonumber \\
&+& \left[ \phi^2(\mathbf{r})
\langle a_0^\dagger a_0 \, \delta \psi^\dagger(\mathbf{r'})\delta \psi(\mathbf{r'})     \rangle + 
\mathbf{r} \leftrightarrow \mathbf{r'} \right]
\label{eq:gIIid}
\\
g_{III}^{(2)} &=&
\langle \delta \psi^\dagger(\mathbf{r})\delta \psi^\dagger(\mathbf{r'})
\delta \psi(\mathbf{r'})\delta \psi(\mathbf{r}) \rangle \,. 
\end{eqnarray}
Averages involving different numbers of operators
$a_0$ and $a_0^\dagger$ vanish since 
the system is in a statistical mixture of Fock states in the harmonic 
oscillator eigenbasis.

The term $g_I^{(2)}$ originates from the condensate mode only. 
Its contribution to $\mbox{Var}(N_L-N_R)$ is zero for symmetry reasons. 
This is a crucial advantage, because it makes our observable
immune to the non-physical fluctuations of the number 
of condensate particles in the grand canonical ensemble, and legitimates
the use of that ensemble.
For the same symmetry reasons, the second line of $g_{II}^{(2)}$ has a zero contribution
to $\mbox{Var}(N_L-N_R)$.
The term $g_{III}^{(2)}$ originates from the non-condensed gas only. 
Below $T_c$ this gas is saturated ($\tilde{z}\simeq 1$) 
to a number of particles scaling as $T^3$, see (\ref{eq:lt_nat}), 
and its contribution to $\mbox{Var}(N_L-N_R)$, maximal at $T=T_c$,
makes the fluctuations in the particle number difference 
only weakly super-poissonian, as already discussed.

Below $T_c$, the first line of $g_{II}^{(2)}$ is thus the important term. 
It originates from a beating between the
condensate and the non-condensed fields. Its contribution
to $\mbox{Var}(N_L-N_R)$ can be evaluated from Wick's theorem \cite{note}:
\begin{multline} 
\label{eq:contII}
\mbox{Var}_{II}(N_L-N_R) = 4 \langle N_0 \rangle 
\left[\int_{\mathbf{r}\in L} \int_{\mathbf{r'} \in L}
\phi(\mathbf{r})\phi(\mathbf{r'}) g^{(1)}(\mathbf{r},\mathbf{r'}) 
\right.
\\
\left.
-\int_{\mathbf{r}\in L} \int_{\mathbf{r'} \in R}
\phi(\mathbf{r})\phi(\mathbf{r'}) g^{(1)}(\mathbf{r},\mathbf{r'}) 
\right] 
\,. 
\end{multline}
Using Eq. (\ref{eq:g1}) and setting $\tilde{z}\simeq 1$ in $g^{(1)}$, after some algebra, we obtain for $T<T_c$
\begin{equation}
\mbox{Var}_{II}(N_L-N_R) \simeq \langle N_0\rangle c_\infty. \label{eq:beat}
\end{equation}
We can thus give a physical meaning to the mathematical splitting (\ref{eq:split}) for $T<T_c$:
The second term and the sum over $s$ in the right-hand side of (\ref{eq:split})
respectively correspond to the condensate-non-condensed beating contribution
$\mbox{Var}_{II}(N_L-N_R)$  and to the purely non-condensed
contribution $\mbox{Var}_{III}(N_L-N_R)$.

\section{Classical field approximation and interacting case}
In this section we show for the ideal gas that the classical field approximation \cite{Kagan,Sachdev,Rzazewski0,Burnett}
exactly gives the high temperature limit ($k_BT \gg \hbar \omega_x$) of the 
amplitude $c_\infty$ in the condensate-non-condensed beating term of $\mbox{Var}(N_L-N_R)$ (\ref{eq:beat}). We then use the classical field
approximation to extend our analysis to the interacting case.

\subsection{Ideal gas: test of the classical field approximation}
It is useful to rewrite (\ref{eq:contII}) as an integral over the whole space introducing the sign function $s(x)$. One then recognizes two closure
relations on $\mathbf{r}$ and $\mathbf{r'}$ and obtains:
\begin{equation}
\label{eq:cinfcomp}
c_\infty = 2 \langle \phi | s(x) \, \frac{1}{z^{-1} e^{\beta h_1}-1} \,  s(x)|\phi \rangle \,. 
\end{equation}
Correspondingly, in the classical field limit: 
\begin{equation}
c_\infty^{\rm class} = 2 \langle \phi | s(x) \, \frac{k_BT }{h_1-\sum_\alpha \frac{\hbar \omega_\alpha}{2}} \,  s(x)|\phi \rangle \,.
\label{eq:cinf_class_0}
\end{equation}
Inserting a closure relation on the eigenstates of the harmonic oscillator $|n\rangle$ in (\ref{eq:cinfcomp}), we are then led to calculate the matrix elements
in one dimension
\begin{equation}
{}_x\langle 0| s(x)  |n\rangle_x = \left( \frac{2\hbar} {m \omega_xn}\right)^{1/2} \phi_{0}^x(0)  \phi_{n-1}^x(0) \,.
\end{equation}
To obtain this result, we introduced the raising operator $a_x^\dagger$ of
the harmonic oscillator along $x$ and we evaluated the matrix element ${}_x \langle 0| [s(x),a_x^\dagger ] |n-1\rangle_x$
in two different ways. First, it is equal to $n^{1/2} {}_x \langle 0 |s(x)|n\rangle_x$
since $a_x^\dagger |n-1\rangle_x = n^{1/2} |n\rangle_x$. Second it can be
deduced from the commutator $[Y(x),a_x^\dagger]=[\hbar/(2m\omega_x)]^{1/2} \delta(x)$
where $Y(x)$ is the Heaviside distribution.
From the known values of $\phi_n^x(0)$ (see e.g. \cite{VarennaYvan}), 
we thus obtain:
\begin{equation}
c_\infty = \frac{4}{\pi} \sum_{m \in \mathbb{N}} \left[ e^{(2m+1)\eta_x}-1\right]^{-1} \frac{(2m)!}{2^{2m}(2m+1)(m!)^2} \,.
\label{eq:clim_occ}
\end{equation}
The equivalent for the classical field is readily computed and we obtain:
\begin{equation}
c_\infty^{\rm class} = \frac{2 \ln 2}{\eta_x}
\end{equation}
showing that the classical field approximation gives the right answer for the dominant contribution to our observable. 
Moreover, going to the first order beyond the classical field approximation, that is
including the $-1/2$ term in the expansion $1/[\exp(u)-1]=u^{-1} -1/2 + O(u)$,
equation (\ref{eq:cinfcomp}) readily gives the term $-1$ in (\ref{eq:euler}). In what follows we will use the classical field approximation
to treat the interacting case.

\subsection{Classical field simulations for the interacting gas}
\begin{figure}[htb]
\centerline{\includegraphics[width=7cm,clip=]{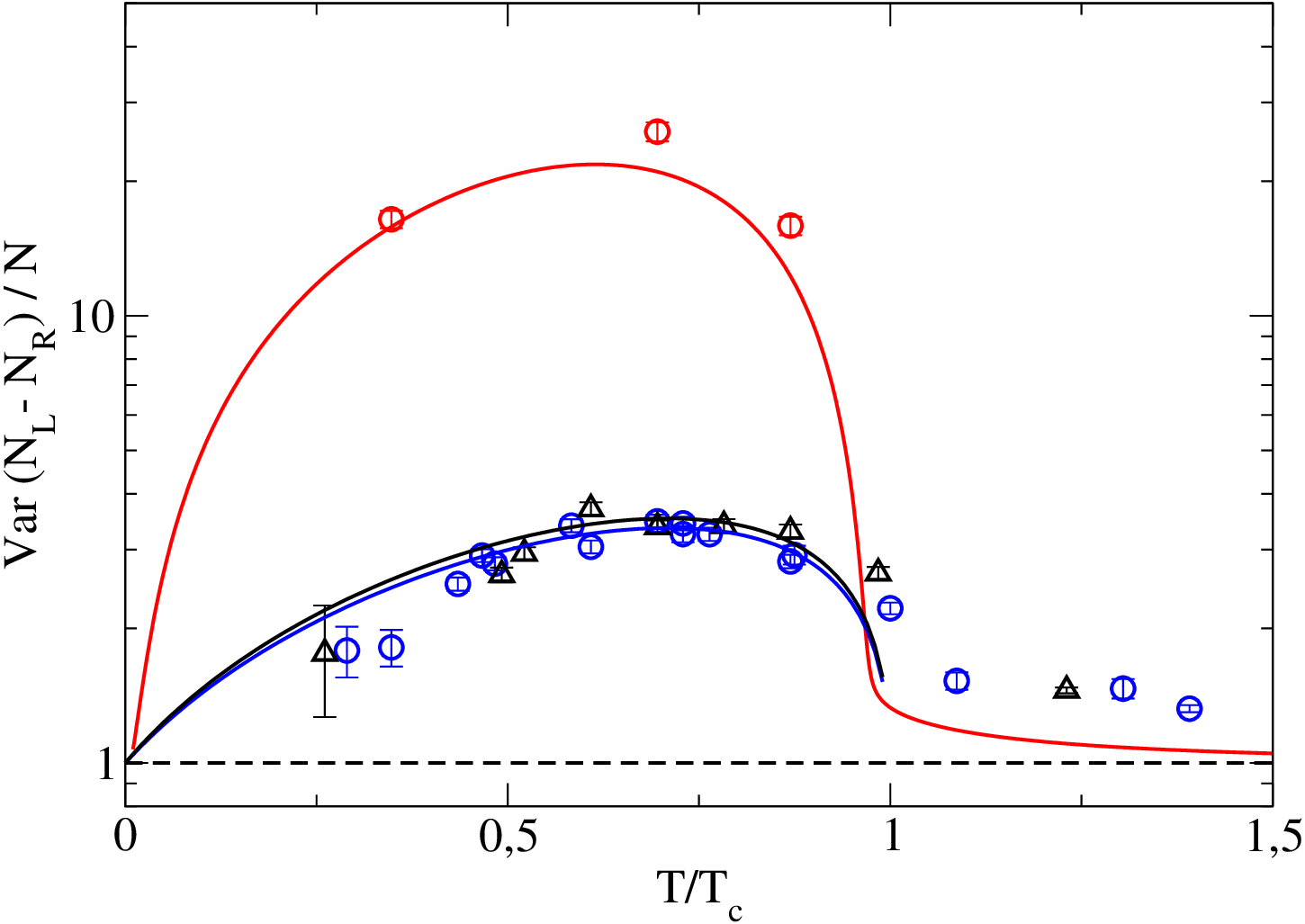}}
\caption{(Color online) Symbols: Classical field simulations for $N=17000$ (circles) and $N=6000$ (triangles) ${}^{87}$Rb atoms with interactions (blue and black),
and without interactions for comparison (red). Red solid line: analytical prediction (\ref{eq:result}) in the non interacting case.  
Blue and black solid lines: analytical prediction (\ref{eq:analy_int}) in the interacting case. The oscillation
frequencies along $x,y,z$ are $\omega_\alpha /2\pi [{\rm Hz}]=(234,1120,1473)$. For the 
interacting case the $s$-wave scattering length is $a=100.4$ Bohr radii.
\label{fig:int} }
\end{figure}

In Fig.\ref{fig:int} we show
results of a classical field simulation in presence of interactions  for two different atom numbers (blue circles and black triangles). 
The non interacting case for one atom number (red circles and red curve) is shown for comparison. 
We note that the assumption of an ideal Bose gas is nowadays realistic:
Recently, the use of a Feshbach resonance has allowed to reach a scattering length
of $a=0.06$ Bohr radii \cite{Inguscio}. We estimate from the Gross-Pitaevskii equation for a pure condensate
that interactions are negligible if $\frac{1}{2} N g \int |\phi|^4 \ll \hbar \omega_{\rm min}$ where
$\omega_{\rm min}$ is the smallest of the three oscillation frequencies $\omega_\alpha$. 
For the parameters of Fig.~\ref{fig:int} (red curve) this results in the well-satisfied condition $N \ll 10^5$.

In presence of repulsive interactions,
the peak of fluctuations in the particle number difference at $T<T_c$ is still present, approximately in the same position, but
its amplitude is strongly suppressed with respect to the ideal gas case. Another notable effect is that the dependence of the curve on the atom number
is almost suppressed in the interacting case.

To compute the normally ordered 
contribution in (\ref{eq:prima}) we generate 800 stochastic fields in the canonical ensemble sampling the Glauber-P function that we approximate by the classical distribution $P\propto \delta\left( N - \int |\psi|^2 \right) \exp\{-\beta E[\psi,\psi^*]\}$ where $E[\psi,\psi^*]$ is the Gross-Pitaevskii energy functional
\be
E[\psi,\psi^*]=\int  \psi^* h_1 \psi + \frac{g}{2} |\psi|^4  \,,
\ee
where the coupling constant $g=4\pi \hbar^2 a/m$ is proportional to the $s$-wave scattering length $a$.
The approximate Glauber-P function is sampled by a brownian motion simulation in imaginary time \cite{Mandonnet}.
The simulation results are plotted as a function of
$T/T_c^{\rm class}$ where the transition temperature in the classical field simulations, extracted by diagonalization
of the one-body density matrix, slightly differs from $T_c$ given by (\ref{eq:Tc}): $T_c^{\rm class} = 1.15 \, T_c$.

\subsection{Analytical treatment for the interacting gas}
In the interacting case, we perform the
same splitting as in (\ref{eq:sepa}) except that now the condensate field $\psi_0=\langle N_0 \rangle^{1/2} \phi$ solves the Gross-Pitaevskii equation
\be
(h_1 + g \psi_0^2) \psi_0 = \mu \psi_0 \,. \label{eq:gpe}
\ee
For the dominant contribution of $g^{(2)}$ to the signal, for $T<T_c$, we then obtain
\bea
g_{II}^{(2)}(\mathbf{r},\mathbf{r'})&=& 2 \langle N_0 \rangle \phi(\mathbf{r})\phi(\mathbf{r'}) \left[
\langle \Lambda^\dagger(\mathbf{r'})\Lambda(\mathbf{r}) \rangle + 
\langle \Lambda(\mathbf{r'})\Lambda(\mathbf{r}) \rangle \right] \nonumber \\
&+& \langle N_0 \rangle \left[ \phi^2(\mathbf{r}) \langle \Lambda^\dagger(\mathbf{r'})\Lambda(\mathbf{r'}) \rangle  +
\mathbf{r} \leftrightarrow \mathbf{r'} \right] \label{eq:g2int} \,,
\eea
where we have neglected fluctuations of $N_0$, set $a_0 = \langle N_0 \rangle^{1/2} e^{i\theta}$ and introduced
$\Lambda(\mathbf{r}) = e^{-i\theta} \delta \psi(\mathbf{r})$ \cite{CastinDum}.
The second line in (\ref{eq:g2int}) brings no contribution to the signal for symmetry reasons. Note that the terms in $g^{(2)}$
that are cubic in the condensate field vanish since $\langle \Lambda(\mathbf{r}) \rangle=0$.

In the Bogoliubov \cite{TestQMC} and classical field approximation, the non-condensed field $\Lambda$ has an equilibrium probability distribution 
\be
P(\Lambda,\Lambda^\ast) \propto \exp \left\{-\frac{\beta}{2} \int (\Lambda^*, \Lambda) \eta \mathcal{L} \left( \begin{array}{l} \Lambda \\ \Lambda^* \end{array}\right)
\right\}
\ee
where the matrix $\eta \mathcal{L}$ is given in \cite{cartago}. Splitting $\Lambda$ into its real and imaginary parts $\Lambda_R$ and $\Lambda_I$, which turn
out to be independent random variables, we then obtain the probability distribution for $\Lambda_R$:
\be
P(\Lambda_R) \propto \exp \left\{-\beta \int \Lambda_R \mathcal{H} \Lambda_R \right\} 
\ee
with
\be
\mathcal{H} = h_1 + 3 g \psi_0^2 -\mu \,.
\ee
The relevant part of $g_{II}^{(2)}$, first line in (\ref{eq:g2int}), can then be expressed in terms  of $\Lambda_R$ only. Passing through the eigenstates
of $\mathcal{H}$ and proceeding as we have done to obtain (\ref{eq:cinfcomp})-(\ref{eq:cinf_class_0}), we then obtain
\be
c_\infty^{\rm class} = 2 k_BT \langle \phi |s(x) \frac{1}{\mathcal{H}} s(x)|\phi \rangle \label{eq:cinf_class_int}
\ee
that is the equivalent of (\ref{eq:cinf_class_0}) for the interacting gas. 
We then have to solve the equation
\be
\mathcal{H} \chi = s(x) \phi \,. \label{eq:surchi}
\ee 
In the absence of the sign function the solution $\phi_a$ of (\ref{eq:surchi}) is known \cite{Lewenstein,CastinDum}: 
\be
\phi_a(\mathbf{r})= \frac{\partial_{\langle N_0 \rangle} \psi_0(\mathbf{r}) }{{\langle N_0 \rangle}^{1/2} \mu'(\langle N_0 \rangle)}
\ee
where $\mu'$ is the derivative of the chemical potential with respect to $\langle N_0 \rangle$. This can be obtained by taking the derivative
of the Gross-Pitaevskii equation (\ref{eq:gpe}) with respect to $\langle N_0 \rangle$. 

In the presence of $s(x)$, the spatially homogeneous case may be solved exactly. One finds that the solution $\chi$ differs from $s(x) \phi_a$
only in a layer around the plane $x=0$ of width given by the healing length $\xi$. In the trapped case, in the Thomas-Fermi limit
where the radius of the condensate is much larger than $\xi$, we reach the same conclusion and we
use the separation of length scales to calculate $\chi$ approximately.
Setting
\be
\chi(\mathbf{r}) = f(\mathbf{r}) \phi_a(\mathbf{r})
\ee
we obtain the still exact equation
\be
-\frac{\hbar^2}{2m} \frac{\phi_a}{\phi} \Delta f - \frac{\hbar^2}{m} \frac{{\bf grad}\,\phi_a}{\phi}\cdot {\bf grad}\, f + f = s(x).
\ee
We expect that $f$ varies rapidly, that is over a length scale $\xi$, in the direction $x$ only, so that we take
$\Delta f\simeq \partial_x^2 f$ and we neglect the term involving $ {\bf grad}\, f$. Also $f$ deviates
significantly from $s(x)$ only at a distance $\lesssim \xi$ from $x=0$, so that $\phi_a/\phi$ may be evaluated
in $x=0$ only. This leads to the approximate equation
\be
-\frac{1}{\kappa^2(y,z)} \partial_x^2 f + f = s(x),
\label{eq:f}
\ee
with
\be
\frac{\hbar^2 \kappa^2(y,z)}{2 m} = \frac{\phi(0,y,z)}{\phi_a(0,y,z)} \simeq 2(\mu - U)
\ee
where $U$ is the trapping potential and the approximation of $\kappa$ was obtained in the Thomas-Fermi approximation $g \psi_0^2 \simeq \mu - U$.
Equation (\ref{eq:f}) may be integrated to give
\be
\chi(\mathbf{r}) \simeq s(x) \phi_a(\mathbf{r}) \left[1-e^{-\kappa(y,z)|x|}\right].
\ee

Neglecting the deviation of $\chi$ from $s(x) \phi_a$ in the thin layer around $x=0$, we finally set
\be
\chi(\mathbf{r}) \simeq s(x) \phi_a(\mathbf{r})
\ee
and obtain the simple result
\be
c_\infty^{\rm clas} \simeq  \frac{k_B T}{\langle N_0\rangle \mu'(\langle N_0\rangle)}.
\ee
For harmonic trapping where $\mu$ scales as $\langle N_0\rangle^{2/5}$, and assuming that $\langle N_0\rangle$
is well approximated by the ideal gas formula (\ref{eq:lt_nat}), we then obtain:
\be
\label{eq:analy_int}
\mbox{Var}\, (N_L - N_R) \simeq \langle N\rangle \left[1 +  \frac{k_B T_c}{\frac{2}{5} \mu(\langle N\rangle)}
\left(1-\frac{T^3}{T_c^3}\right)^{3/5} \frac{T}{T_c}
\right].
\ee
The analytic prediction (\ref{eq:analy_int}) is plotted as a full line (black and blue for two different atom numbers) in Fig.\ref{fig:int}.
We note a good agreement with the numerical simulation. From (\ref{eq:analy_int}) we can extract the position of the maximum in the
normalized fluctuations:
\be
\left(\frac{T}{T_c}\right)_{\rm max} = \frac{980^{1/3}}{14} \simeq 0.709
\ee
as well as their amplitude:
\be
[\mbox{Var}\, (N_L - N_R)]_{\rm max} \simeq \langle N\rangle \left[1+ 1.36 \frac{k_B T_c}{ \mu(\langle N\rangle)}\right].
\label{eq:result_int}
\ee
Note the very weak dependence of $k_B T_c/\mu(\langle N\rangle)$ on the atom number, scaling as $\langle N\rangle^{-1/15}$
\cite{hydro}. We point out that our analysis in the interacting case is quite general and can be applied to atoms in any even trapping potential
provided that the potential does not introduce a length scale smaller than the healing length $\xi$. 

\subsection{First quantum corrections to the classical field}
\label{sec:quantcorr}
Expanding the non-condensed fields $\Lambda(\rr)$ and $\Lambda^\dagger(\rr)$ over Bogoliubov modes:
\begin{equation}
\left(\begin{array}{c}{\Lambda(\rr)} \\ {\Lambda}^\dagger(\rr)\end{array}\right)=
\sum_{j} \:
b_j \left(\begin{array}{c}{u_j(\rr)} \\ {v_j(\rr)}\end{array}\right)+
b_j^\dagger \left(\begin{array}{c}{v_j^*(\rr)} \\ {u_j^*(\rr)}\end{array}\right)
\label{eq:dev}
\end{equation}
and using (\ref{eq:g2int}), the coefficient $c_\infty$ giving the dominant contribution (\ref{eq:beat})
to the signal for $T<T_c$,  can be split as 
$c_\infty=c_\infty^{\rm th}+c_\infty^{0}$. With the thermal contribution
\be
c_\infty^{\rm th}= 2 \sum_j \bar{n}_j |\langle \phi | s(x) (|u_j \rangle + |v_j \rangle)|^2 \geq 0 \label{eq:cinfth_uv}
\ee  
where $\bar{n}_j=\langle b_j^\dagger b_j \rangle =1/[\exp(\beta \epsilon_j) -1]$, $\epsilon_j$ being the energy of the Bogoliubov mode $j$,
and the zero temperature contribution
\be
c_\infty^{0}= 2 \sum_j  |\langle \phi | s(x) |v_j \rangle|^2 + \langle \phi |s(x)| u_j \rangle \langle v_j |s(x) |\phi \rangle \,.
\ee  
Performing the classical field approximation that is setting $\bar{n}_j=k_B T/\epsilon_j$ in (\ref{eq:cinfth_uv}), exactly gives the expression
(\ref{eq:cinf_class_int}) of $c_\infty^{\rm class}$ \cite{equivalence}. Introducing the first quantum correction $-1/2$ to the occupation number $\bar{n}_j$ and
including the quantum contribution, we obtain the first quantum correction to $c_\infty^{\rm class}$ in the interacting case $\delta c_\infty = c_\infty^0 + \delta c_\infty^{\rm th}$:
\be
\delta c_\infty =  \sum_j \langle \phi | s(x) |v_j \rangle \langle v_j | s(x) |\phi \rangle -
\langle \phi | s(x) |u_j \rangle \langle u_j | s(x) |\phi \rangle \,.
\ee
Using the closure relation $\sum_j |u_j\rangle \langle u_j| - |v_j\rangle \langle v_j| = 1-|\phi \rangle \langle \phi|$ \cite{CastinDum}, we obtain
\be
\delta c_\infty = -1 \,.
\ee  
This exactly corresponds to the first correction in (\ref{eq:euler}) for the ideal gas. We conclude that also in the interacting case,
the first quantum correction to $c_\infty^{\rm class}$ has the effect of changing the shot noise term (the first term equal to one) in the square brackets of (\ref{eq:analy_int}) into
$1-\langle N_0\rangle/\langle N \rangle$. 

\section{Conclusion}
We have studied fluctuations in the difference of number of particles in the left ($x<0$) and right ($x>0$) halves of a three dimensional harmonically
trapped Bose gas 
as a function of the temperature across the critical temperature for condensation. 
Both for the ideal gas and the interacting gas, fluctuations are weakly super poissonian for $T>T_c$. If one lowers the temperature from $T_c$
down to zero, fluctuations increase, 
reach a maximum and then decrease again as the non-condensed fraction vanishes. 
We have solved this problem analytically for the ideal gas case, and we have found an 
approximate solution in the interacting case in the Thomas-Fermi limit when the temperature is larger than the quantum of oscillation in the trap.
Remarkably, the local density approximation fails for this problem for the ideal gas. On the contrary, we show that the classical field approximation
correctly gives the high temperature contribution to the fluctuations in the particle number difference both for the ideal gas and the interacting gas.

For a large ideal gas, the maximum of normalized fluctuations $\mbox{Var}(N_L-N_R)/\langle N\rangle$ is located at $T/T_c=2^{-2/3}$ independently
on the trap oscillation frequencies and its amplitude approximately scales as $N^{1/3}\bar{\omega}/\omega_x$. 
For the interacting case in the Thomas-Fermi regime 
the maximum of fluctuations in the relative number of particle subsists approximately for the same value of $T/T_c$ 
but its amplitude is strongly reduced as well as its dependence on $N$ scaling as $N^{-1/15}$.  

\begin{figure}[t!b]
\centerline{\includegraphics[width=7cm,clip=]{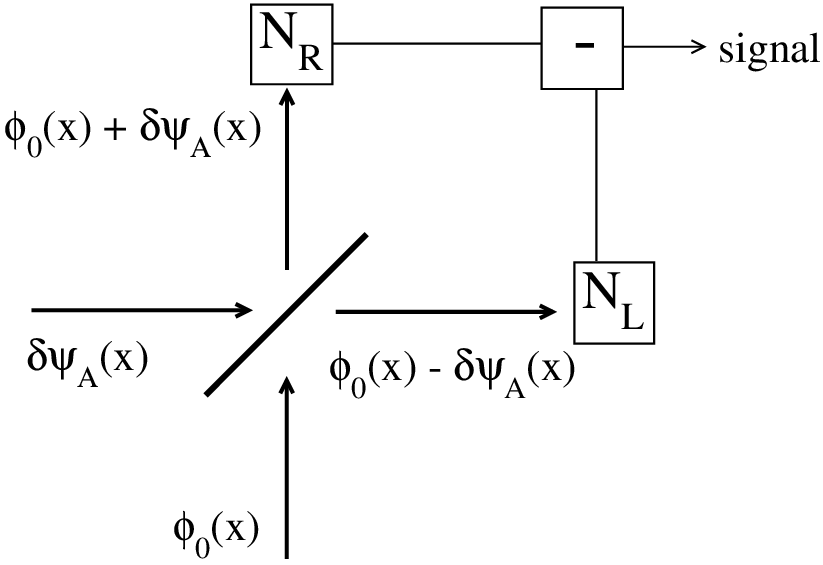}}
\caption{The measurement can be seen
as a balanced homodyne detection of the non-condensed field where the condensate acts as a local oscillator \label{fig:homodyne} }
\end{figure}

Finally, we give a physical interpretation of the ``bump" in fluctuations for $T<T_c$ as due to
a beating between the condensate mode $\phi_0(\mathbf{r})$ and the non-condensed modes $\delta \psi(\mathbf{r})$,
as it is apparent in Eqs.~(\ref{eq:gIIid},\ref{eq:g2int}).
For symmetry reasons, only the antisymmetric component of the non-condensed
field $\delta \psi_A(x)=-\delta \psi_A(-x)$ contributes to $\mbox{Var}(N_L-N_R)$.
The measurement can then be seen in a pictorial way
as a balanced homodyne detection of the non-condensed field where the condensate field acts as a local oscillator (Fig.\ref{fig:homodyne}). 
Since one or the other of these two fields vanishes for $T$ tending to $0$ or $T_c$, the beating effect,
and thus $\mbox{Var}(N_L-N_R)$, are obviously maximal at some intermediate temperature.

\acknowledgments
We thank J. Reichel, J. Est\`eve, K. Maussang and R. Long for stimulating discussions. The teams of A.S. and Y.C. are parts of
IFRAF. L. Y. acknowledges financial support from the National Basic Research Program of China (973 Program) under Grant No.
2006CB921104.


\begin{thebibliography}{99}
\bibitem{Molmer}M. Budde, K. M\o lmer,  Phys. Rev. A {\bf 70}, 053618 (2004).
\bibitem{Houches2003} Y. Castin, ``Simple theoretical tools for low dimension Bose gases",
Lecture notes of the 2003 Les Houches Spring School, Quantum Gases
in Low Dimensions, M. Olshanii, H. Perrin, L. Pricoupenko, Eds., J.
Phys. IV France {\bf 116}, 89 (2004).
\bibitem{VarennaYvan}
 Y. Castin, ``Basic tools for degenerate Fermi gases",
Lecture notes of the 2006 Varenna Enrico Fermi School on
Fermi gases,
M. Inguscio, W. Ketterle, C. Salomon, Eds.,  SIF (2007).
\bibitem{Combescot}
G.E. Astrakharchik, R. Combescot, L.P. Pitaevskii,
Phys. Rev. A {\bf 76}, 063616 (2007).
\bibitem{Brustein}
R. Brustein, A. Yarom, Journ. of Stat. Mech., P07025 (2008).
\bibitem{Levitov}  L.S. Levitov, H.-W. Lee, and G.B. Lesovik, J. Math. Phys. {\bf 37}, 10 (1996).
\bibitem{Bouchoule} J. Esteve, J.-B. Trebbia, T. Schumm, A. Aspect, C.I. Westbrook, I. Bouchoule,
Phys. Rev. Lett. {\bf 96}, 130403 (2006).
\bibitem{Kenneth} K. Maussang et al. {\it in preparation}; Presentation of J. Reichel at the BEC 2009 conference, San Feliu de Guixols, Spain.
\bibitem{LesHouches} Y. Castin, p.1-136, in Coherent atomic matter waves, Lecture notes of
Les Houches summer school, edited by R. Kaiser, C. Westbrook, and
F. David, EDP Sciences and Springer-Verlag (2001).
\bibitem{Landau} L. Landau, E. Lifchitz, p. 103-104, \S 30 
in {\it Physique th\'eorique, Tome V : Physique Statistique}, 
4th edition, Mir (1994).
\bibitem{Integrales} For $f(x,x')=\exp [-\alpha(x+x')^2-\beta (x-x')^2]$ and for $\beta,\alpha>0$, let
$I \equiv \int_{-\infty}^0 dx \int_{-\infty}^0 dx' f(x,x')-\int_{-\infty}^0 dx \int_{0}^\infty dx' f(x,x') \,.$
Using polar coordinates we obtain
$I=\frac{1}{2\sqrt{\alpha \beta}}\arctan \frac{\beta - \alpha}{2\sqrt{\alpha \beta}} \,. \:\:\:\: \mbox{For} \:\:\: \beta > \alpha \:\:\: \mbox{this is also}$
$I=\frac{1}{\sqrt{\alpha \beta}}\left[\frac{\pi}{4} - \frac{1}{2} \arctan \frac{2\sqrt{\alpha \beta}}{\beta-\alpha}\right]=
\frac{1}{\sqrt{\alpha \beta}}\left[\frac{\pi}{4} - \arctan \sqrt{\frac{\alpha}{\beta}}\right] \,.$
\bibitem{lda} Using (\ref{eq:contII}) and  (\ref{eq:beat}), with the local density approximation (\ref{eq:g1lda}) for $g^{(1)}$ (with $\tilde{z}=1$), one obtains
$
c_\infty^{LDA}=\frac{4}{\pi} \sum_{l=1}^\infty \frac{\arctan [\frac{1}{2} \left( \frac{l \eta_x}{2} \right)^{-1/2} \left(1-  \frac{l \eta_x}{2} \right) ]}{
\Pi_\alpha \left( 1 +  \frac{l \eta_\alpha}{2}\right)} \,.
$
This result is qualitatively incorrect: It depends on $\omega_y$ and $\omega_z$ whereas the exact expression (\ref{eq:clim}) does not.
It is quantitatively incorrect even in the isotropic case, where $c_\infty^{\rm LDA} \sim (1/2)/\eta_x$ for $\eta_x\to 0$,
to be compared to (\ref{eq:climint}).
\bibitem{Qui_citer} S. Grossmann, M. Holthaus, Z. Naturforsch. A {\bf 50}, 921 (1995); S. Grossmann, M. Holthaus,
Phys. Lett. A {\bf 208}, 188 (1995).
\bibitem{details}
From the expansion
$f(x)=2-(4\sqrt{2}/\pi) (\eta_x x)^{1/2}+O(\eta_x x)^{3/2}$,
the $k$-th order derivative of $f$ 
in $x=1$ is
$f^{(k)}(1) \simeq -(2/\pi)^{3/2} \eta_x^{1/2} 
(-1)^{k+1}\Gamma(k-1/2)$,
and the integral $\int_0^1 f(x)$ is readily evaluated 
to leading order in $\eta_x$. This leads to (\ref{eq:euler}),
with the numerical coefficient $A=(2/\pi)^{3/2} [S+\pi^{1/2}/3]$, 
where $S$ is a sum involving Bernoulli numbers $B_{2k}$,
$S=\sum_{k\geq 1} \Gamma(2k-3/2) B_{2k}/(2k)!\simeq 0.1461$.
On the other hand, using (i) the integral representation of the Riemann $\zeta$ function, see \S 9.512 with $q=1$ in \cite{Gradstein},
(ii) the definition of Bernoulli numbers \S 9.610 in \cite{Gradstein}, and (iii) the definition of the $\Gamma$ function \S 8.310(2)  in \cite{Gradstein},
we obtain $\zeta(z)= \pi^{-1} \Gamma(1-z) \sin(\pi z) \sum_{n \in \mathbb{N}} (-1)^n \Gamma(z+n-1) B_n / n!$, $\forall z \in \mathbb{C} \setminus \mathbb{Z}$.
Then the desired identity follows from the fact that $B_0=1$, $B_1=-1/2$,  $B_{2k+1}=0 \:  \forall k \geq 1$. 
An alternative technique is to obtain from (\ref{eq:clim_occ}) the integral representation: 
$A=(2/\pi)^{3/2} \int_0^{+\infty} dx \, x^{-3/2} \left[ (e^x-1)^{-1} - x^{-1} + 1/2 \right].$
Using the integral representation of the Riemann $\zeta$ function, see \S 9.512 with $q=1$ in \cite{Gradstein}, we obtain $A = - 2^{5/2} \zeta(-1/2)/\pi $.
\bibitem{note}
In principle in (\ref{eq:contII}) one should have introduced
the first-order coherence function of the non-condensed
field $g^{(1)}(\mathbf{r},\mathbf{r'}) - \langle N_0\rangle
\phi(\mathbf{r})\phi(\mathbf{r'})$, but this does not affect
the result due to symmetry reasons.
\bibitem{Kagan}
Yu. Kagan, B.V. Svistunov, and G.V. Shlyapnikov,
Sov. Phys. JETP {\bf 75}, 387 (1992);
 Yu. Kagan and B. Svistunov, Phys. Rev. Lett. {\bf 79} 3331
(1997).
\bibitem{Sachdev}
K. Damle, S.N. Majumdar and S. Sachdev, Phys. Rev. A {\bf 54}, 5037 (1996).
\bibitem{Inguscio}
M. Fattori, C. D'Errico, G. Roati, M. Zaccanti, M. Jona-Lasinio, M. Modugno, M. Inguscio,
G. Modugno, Phys. Rev. Lett. {\bf 100}, 080405 (2008).
\bibitem{Rzazewski0}
K. G\'oral, M. Gajda, K. Rz\c{a}\.zewski, Opt. Express {\bf 8}, 92 (2001);
D. Kadio, M. Gajda and K. Rz\c{a}\.zewski, Phys. Rev. A {\bf 72}, 013607 (2005).
\bibitem{Burnett}
M.J. Davis, S.A. Morgan and K. Burnett, Phys. Rev. Lett. \textbf{87}, 160402 (2001).
\bibitem{Mandonnet} E. Mandonnet PhD thesis University Paris 6, URL http://tel.archives-ouvertes.fr/tel-00011872/fr/ .  
\bibitem{CastinDum} Y. Castin, R. Dum, Phys. Rev. A {\bf 57}, 3008 (1998).
\bibitem{TestQMC} The accuracy of the Bogoliubov approximation to calculate the pair distribution function
was checked by exact Quantum Monte Carlo calculations for a trapped gas \cite{Holzmann}.
\bibitem{Holzmann} M. Holzmann, Y. Castin, Eur. Phys. J. D {\bf 7}, 425 (1999).
\bibitem{cartago} A. Sinatra, C. Lobo, Y. Castin, J. Phys. B {\bf 35}, 3599 (2002).
\bibitem{Lewenstein}P. Villain, M. Lewenstein, R. Dum, Y. Castin, L. You, A. Imamoglu,
T.A.B. Kennedy, Journal of Modern Optics, {\bf 44}, 1775 (1997).
\bibitem{hydro}
Using the formulation in terms of Bogoliubov modes presented in section \ref{sec:quantcorr}, one can show that the scaling $c_\infty^{\rm class}\propto k_BT/\mu$
for $k_BT \gg \hbar \omega_\alpha$  can also be obtained from an hydrodynamic approximation
to the Bogoliubov mode functions $u_j$ and $v_j$. Introducing the hydrodynamical variables $\rho$ and $S$ such that the time-dependent condensate
field $\psi=\rho^{1/2} \exp(iS/\hbar)$, one linearizes the time-dependent Gross-Pitaevskii equation around the stationary solution $\psi_0$. Differentiating 
$\rho=\psi^\ast \psi$ and  $\psi/\psi^\ast=\exp(2iS/\hbar)$, and setting $\delta \psi=u_j$ and $\delta \psi^\ast=v_j$, we obtain
$
\delta \rho_j= \psi_0 (u_j+v_j) \hspace{0.25cm} {\rm and} \hspace{0.25cm} \delta S/\hbar=(u_j-v_j)/(2i \psi_0) \,.
$
For real mode functions the normalization condition $\int u_j^2 -v_j^2 =1$ together with Euler's equation 
$-i \epsilon_j \delta S_j/\hbar=-g \delta \rho_j$ gives the normalization condition
$
 \frac{2 g}{\epsilon_j} \int_{U({\mathbf {r}})<\mu} d^3r \, \delta \rho_j^2 =1 \,.
$
Let us consider for simplicity the isotropic case (see e.g. \cite{Sinha} for the general anisotropic case).
In the hydrodynamic approximation, $\delta \rho$ is a product of a polynomial $P(r/R)$ where $R$ is the Thomas-Fermi radius, a spherical harmonic and
a normalization factor ${\cal N}_j$. The coefficients of the polynomial are numbers that were calculated in \cite{StringariPRL}. We then find that 
${\cal N}_j^2$ scales as $\epsilon_j/(gR^3)$ so that 
the matrix element squared $|\langle \phi | s(x) (|u_j \rangle + |v_j \rangle)|^2$ scales as $\epsilon_j/\mu$. On the other hand $\bar{n}_j\simeq k_BT/\epsilon_j$
for the low frequency modes. We then find the scaling $c_\infty^{\rm class} \propto k_BT/\mu$.
\bibitem{Gradstein} I.S. Gradshteyn and I.M. Ryzhik, {\it Table of integrals, Series, and Products}, 5th edition,  Academic Press (San Diego, 1994).
\bibitem{Sinha} S. Sinha, Y. Castin, Phys. Rev. Lett. {\bf 87}, 190402 (2001).
\bibitem{StringariPRL} S. Stringari, Phys. Rev. Lett. {\bf 77}, 2360 (1996).
\bibitem{equivalence} This may be checked explicitly by calculating the inverse of $\eta {\cal L}$ from the spectral decomposition
$\left( \eta {\cal L}\right)^{-1}={\cal L}^{-1}\eta=\sum_j \left( 1/\epsilon_j \right) \left[ \left( \begin{array}{c} |u_j\rangle \\ | v_j \rangle 
\end{array} \right) \left( \langle u_j | , \langle v_j | \right) + \left( \begin{array}{c} |v_j^*\rangle \\ | u_j^* \rangle 
\end{array} \right) \left( \langle v_j^* | , \langle u_j^* | \right)  \right]$ with $\eta = \left( \begin{array}{rr} 1 & 0 \\ 0 & -1 \end{array} \right)$,
see e.g. \cite{CastinDum}.
\end{thebibliography}
\end{document}